# On the influence of water on THz vibrational spectral features of molecular crystals

Sergey Mitryukovskiy,[a] Danny E. P. Vanpoucke,[b] Yue Bai,[a] Théo Hannotte,[a] Mélanie Lavancier,[a] Djamila Hourlier,[a] Goedele Roos [c] and Romain Peretti [*a]

a. Institut d'Electronique de Microélectronique et de Nanotechnologie, Université Lille, CNRS, 59652 Villeneuve d'Ascq, France. E-mail: romain.peretti@univ-lille.fr
b. IMO, Hasselt University, 3590 Diepenbeek, Belgium./ AMIBM, Maastricht University, 6167 Geleen, The Netherlands
c. Univ. Lille, CNRS, UMR 8576 - UGSF - Unité de Glycobiologie Structurale et Fonctionnelle, F-59000 Lille, France

**Abstract** – The nanoscale structure of molecular assemblies plays a major role in many (μ)-biological mechanisms. Molecular crystals are one of the most simple of these assemblies and are widely used in a variety of applications from pharmaceuticals and agrochemicals, to nutraceuticals and cosmetics. The collective vibrations in such molecular crystals can be probed using terahertz spectroscopy, providing unique characteristic spectral fingerprints. However, the association of the spectral features to the crystal conformation, crystal phase and its environment is a difficult task. We present a combined computational-experimental study on the incorporation of water in lactose molecular crystals, and show how simulations can be used to associate spectral features in the THz region to crystal conformations and phases. Using periodic DFT simulations of lactose molecular crystals, the role of water in the observed lactose THz spectrum is clarified, presenting both direct and indirect contributions. A specific experimental setup is built to allow the controlled heating and corresponding dehydration of the sample, providing the monitoring of the crystal phase transformation dynamics. Besides the observation that lactose phases and phase transformation appear to be more complex than previously thought – including several crystal forms in a single phase and a non-negligible water content in the so-called anhydrous phase – we draw two main conclusions from this study. Firstly, THz modes are spread over more than one molecule and require periodic computation rather than a gas-phase one. Secondly, hydration water does not only play a perturbative role but also participates in the facilitation of the THz vibrations.

## 1. Introduction

Water molecules play an essential role in the structural and functional properties of organic (and even inorganic) compounds. They can easily be incorporated in organic compounds, hereby creating hydrates through the formation of sub-lattices, organized within the crystal of the host compound. Because water acts as both a hydrogen bond donor and acceptor, it can interact with many hydrogen-bonding moieties in organic systems. It is therefore not surprising that water in all its forms is probably one of the most widely studied chemical compounds[1]. A wide range of experimental techniques can be used to address the role of water in organic molecular structures: X-ray and neutron diffraction[2], nuclear magnetic resonance (NMR)[3,4] and various optical spectroscopic techniques[5,6]. Computational studies (most commonly, Density Functional Theory (DFT) calculations and Molecular Dynamics (MD) simulation) can complement the knowledge gained through experiments and offer a description of the macromolecule or the molecular crystal and solvent as well as their dynamics at the atomic scale[7–11].

From the theoretical point of view, water offers the opportunity to enlarge our understanding of hydrogen bonding[12]. Several recent reviews exist on the role of water in the structure and dynamics of proteins[13], its hydration and ligand recognition[14], DNA[15,16] and other macromolecular structures[17]. In addition, water of crystallization is crucial for food science[18–22] and pharmacology[5,23–25]. Furthermore, water molecules can interact with pharmaceutical compounds in numerous ways, which can dramatically affect the performance of a final dosage form. For both food science and pharmacology, the role of water in mono- and polysaccharides is extremely interesting[26–30]. Overall, understanding the role of water in molecular crystals and macromolecules is still an actual challenge that can be tackled using vibrational spectroscopy.

Vibrational spectroscopy offers a number of experimental techniques operating in a large range of wavelengths, from the visible to microwave spectrum, including near-, mid-, far-infrared (IR) and terahertz (THz) regions. Please note that the terahertz region is equivalent to the far-infrared region, and ranges from 0.1 to 10 THz. In each of these regions, different molecular bonds, delocalized vibrational modes and molecular motions are revealed as the characteristic signatures of the chemical structure. Vibrational spectroscopy is able, not only to identify the composition of molecular crystals, but also to register its transformation dynamics (kinetics, solvent-mediated, hydration, etc.)[31–34]. For example, hydration/dehydration of molecular crystals and saccharides, has been studied by means of Mid-IR[35], Near-IR[36,37] and Raman[38] spectroscopy.

THz technologies progressed tremendously over the past two decades and currently THz spectroscopy and non-destructive testing systems begin to be commercially available. It led to various applications in security[39], medical diagnostics[40], pharmacology[41], automotive[42] and food[43] industries. More precisely, time-domain spectroscopy (TDS) enabled broadband THz absorption spectroscopy[32]. In crystalline solids, electromagnetic waves can be used to probe the phonons propagating through the crystal structure. Within the THz spectral range, the energy is lower than in mid-IR. As a result, the vibrations that are probed are of a much longer length scale than individual molecular bonds, studied using IR spectroscopy.

The fact that THz spectra include information of both water of crystallization and the host crystal structure makes THz-TDS very sensitive to dehydration. This is shown by many studies on the structure and dynamics of bound water in glucose[44] and polymers[45,46], the dehydration of crystalline proteins[47], and the observation of the critical effect of hydration on the resonant signatures of biomolecules[48]. During dehydration, THz spectroscopy will precisely monitor both the loss of water of crystallization and the eventual molecular rearrangement. This could then be further extended to study systems where multiple solid-state forms appear as a result of dehydration.

Because (hydrated) molecular crystals and isomorphous dehydrates consist of hundreds or even thousands of atoms, highly accurate computational studies at the quantum mechanical level are extremely demanding. In addition, the methodological requirements for high quality calculations in molecular crystals are only recently being developed[49,50]. The interpretation of the complex data obtained in THz spectroscopy experiments requires the use of advanced theoretical approaches and robust computational methods[49], due to the intricacy of the low-frequency vibrations and their sensitivity to the structural environment. Calculations making use of density functional theory (DFT), a leading first-principle approach, can yield reliably simulated spectra. It is commonly used to predict gas-phase spectra of isolated molecules, dimers, and clusters[12,51–53], and is becoming more common for solid-state systems as well. However, in solid-state systems this remains a challenging problem due to the complexity involved in periodic DFT modelling of large systems like macromolecules and molecular crystals[50,54,55].

Alternatively, one can obtain the vibrational spectrum, including intermolecular modes through classical molecular dynamics (MD) simulations[56,57] via the analysis of the Fourier transform of the velocity autocorrelation function. Although limited by the approximations of the effective force fields utilized to describe the intermolecular interactions, the lower computational cost provides access to much larger systems than are currently feasible at the DFT level, making MD simulations appealing for the analysis of the THz domain and allowing the detection of specific molecules present in mixtures in condensed phases[7].

Despite the progress in the simulations of the solid-state THz response, the calculation of spectral features remains one of the greatest challenges for computational molecular spectroscopy, as no universal approach exists yet. Many different solid-state DFT codes[10,58–61] implement the DFT formalism in a different way, putting in doubt the reproducibility of such predictions[62]. Furthermore, any computational strategy for the calculation of phonons in the THz regime suffers from convergence issues. As a result, a step-by-step approach is required to ensure the convergence of the results. Specifically, if one begins with a system in mechanical equilibrium, meaning that all atomic forces are (extremely close to) zero, one needs to perform a series expansion in order to calculate phonons. In addition, the Hellman-Feynman forces have to be described with an exceptionally high accuracy for the results to accurately render the shallow and low energy parts of the potential energy surface (PES). An immediate consequence is that the computational cost will increase similarly. From a recent study by Banks *et al.*, it is also clear that the description of the vibrational spectrum of molecular crystalline materials requires taking into account the periodic arrangement of the molecules[49]. Excellent reviews of DFT studies on water (monomer, dimer, clusters, hexamer, ice structures, liquid water) have been presented, such as the one by Gillan *et al.*[12]

In this study, we consider lactose as our example case. The monohydrate-crystallized form of lactose has become the "gold standard" for THz spectroscopy[34,63], due to its strong and narrow absorption peaks, notably at 0.53 and 1.37 THz – a typical absorption spectrum of α-lactose monohydrate[54,64–66] is presented in Figure 1b. Also, due to its wide availability and low cost, lactose samples are often used to test THz spectroscopy equipment[34,67], serving as a "first-try" sample in minor-volume detection techniques[68–70], and is used as a mixture compound in approving methods for content quantification[71,72]. Furthermore, the study of lactose is important for pharmacology[73,74], medicine[75–77] and food industry[78]. Being a polysaccharide, lactose strongly interacts with water molecules forming hydrogen bonds[19]. Lactose is highly polymorphic, with the currently known forms (note that we adopt here the nomenclature given in Y. Listiohadi *et al.*[79]): α-lactose monohydrate, α-lactose anhydrous unstable (also known as hygroscopic α-lactose), α-lactose anhydrous stable (also known as anhydrous α-lactose), β-lactose anhydrous, amorphous lactose and β/α-lactose compound crystals. Throughout the paper, we will use the following terminology: α-lactose monohydrate (α-LM), hygroscopic α-lactose (α-LH) and anhydrous α-lactose (α-LA). The water content varies among these polymorphs and is altered during storage depending on the relative humidity. The initial water content of α-lactose monohydrate is determined to be 5.26%, which can be increased to 6.17% after 3 months of storage at 75% relative humidity. The anhydrous forms do contain some initial water, depending on the polymorph (see Table 2 of [79]) and were found to be highly hygroscopic during storage, even at low relative humidities.

This makes lactose a perfect compound to study the influence of water in molecular crystals on the spectral features. However, the interpretation of the data is complicated by the varying water content. The theoretical explanation of the lactose THz spectrum is rather rare in literature and is mostly based on gas phase DFT calculations considering a single lactose molecule, and not crystals, meaning the collective vibrations are not taken into account[49]. Here, we provide calculations on crystalline lactose polymorphs, giving a better representation of the experimental reality compared to the single-molecule calculations. This is however hampered due to the fact that the experimentally prepared polymorphs contain variable water content, and even the anhydrous forms are not completely water free[79], which cannot be represented by the theoretical calculations. This shows the advantage and drawback of calculations : in calculations, the polymorphs are completely 'pure' forms (for example, the anhydrous forms are completely water-free in theoretical calculations), but unfortunately, this is not the case in experimentally prepared polymorphs[79], giving rise to discrepancies and thus the interpretation of the results is not straightforward.

In this work, the vibrational modes of the gas phase lactose molecule, and their apparent agreement with the experimental ones, is discussed. From these observations, we continue with dehydrated lactose crystals and proceed with the incorporation of water molecules herein. The resulting vibrational spectra are used to elucidate the central role of the water molecules in the molecular crystals. To deepen our understanding of the role of water and to validate the simulation results, an experimental setup was developed. It allows the heating and temperature control of the samples, and thus provides the possibility for the dehydration. We observed the spectral signatures of three different "metastable" crystal phases (stable in $N_2$ atmosphere, namely, the α-lactose monohydrate (α-LM or phase 1), and two other phases, hereafter called phase 2 and 3) in agreement with the predictions by calculations and mass-spectroscopy tests. Our setup also allows monitoring of the transformation dynamics between the phases.

To provide a clearly structured and comprehensive manuscript, we present first the experimental results and then the computational results. We would like to note that the study has started with the theoretical approach and followed by the experimental confirmation of our findings.

## 2. Materials and methods

### 2.1. Lactose samples

α-Lactose Monohydrate ($C_{12}H_{22}O_{11} \cdot H_2O$) (α-LM) powder purchased from Sigma-Aldrich Co. Ltd. (≥ 99.9% total lactose basis including less than 4% β-lactose) was used as the sample material without further purification. The powder was stored under normal conditions (20°C, 1 atm.), in the closed original container. The relatively low humidity of our storage condition (45±5%) ensured the stability of the polymorph[79]. Several pellets were prepared from the lactose powder, with a thickness of ~ 0.6-0.7 mm and the diameter of ~ 10 mm (PIKE CrushIR digital hydraulic press). Although the experiments have been repeated on several pellets confirming the reproducibility, all experimental results presented in this work were obtained with the two similar pellets (the first one for the monohydrate to phase 2 transition, the second – to obtain the phase 3). This allows us to show the dynamics without changing additional parameters (as initial pellet thickness, water content, etc.).

A typical THz absorption spectrum of the α-LM pellet (always stored under normal conditions, but placed into a nitrogen-purged box prior to the experiment) is presented in Figure 1b and agrees well with the widely known α-LM spectra from the literature[64].

### 2.2. Experimental setup

The THz-TDS experiments were performed using the TeraSmart system from MenloSystems Gmbh upgraded with an in-house-made pellet holder with integrated heating and its controller (*cf.*, Figure 1a for the schematics). A set of four 50-mm-focal-length Polymethylpentene ("TPX") lenses served to focus the THz pulses in the sample pellet centre. The THz emitter and detector, the lenses and the pellet holder were placed inside a Plexiglas glove-box purged with constant nitrogen flow (low, but high enough to compensate the leakages of the box and to ensure a slightly higher pressure than the laboratory room one) to prevent the absorption of THz emission by atmospheric water vapour. The setup was kept purging for about 24 hours before performing the measurements, this assures a regular spectral curve as shown in Figure 1b. No data treatment was performed to remove the water features.

The sample pellet is placed inside a mount specially designed for the experiment (Figure 1c) allowing the fixation of the pellet, its heating at a set rate, and controlling its temperature, with the simultaneous THz-TDS measurements. The mount is made of two aluminium parts to hold a pellet sample inside and with a 5-mm-diameter round opening around the pellet centre to allow the propagation of the THz pulses through the pellet. Two 12-V, 40-Watt heating cartridges (typically used in 3D printers) inserted inside the mount assure the heating. The temperature of the pellet is controlled with the T-type thermocouple integrated into the mount, touching (deepen inside) the edge of the pellet. All the elements as well as the pellet surfaces contained inside the mount (not in its opening) are greased with the Arctic MX-2 thermal compound paste to ensure better thermal conductivity. Although the direct measurement of the temperature is done at the edge of the pellet, comparison to the pellet centre temperature in a number of tests allowed us to calibrate for the pellet centre temperature with a precision of ~2 K, sufficient for our study[80]. In the following, we will always refer to the temperature at the pellet's centre. The temperature control is achieved by using the thermocouple and the two cartridges controlled by a solid-state relay connected to a microcontroller board interfacing with LabVIEW based software. The software is able to communicate with the TeraSmart system and allows to command the measurement process and to retrieve the THz-TDS data. The software allows the heating at a constant rate (most convenient for the experiment was ~ 3 K/min) and the temperature stabilization at a fixed value by means of the proportional–integral–derivative (PID) control loop mechanism. Moreover, the experiment presented in this work is completely automated and, with a properly input procedure (temperature steps and number of measurements), can be done in one single round (with the total duration of around 3 hours).

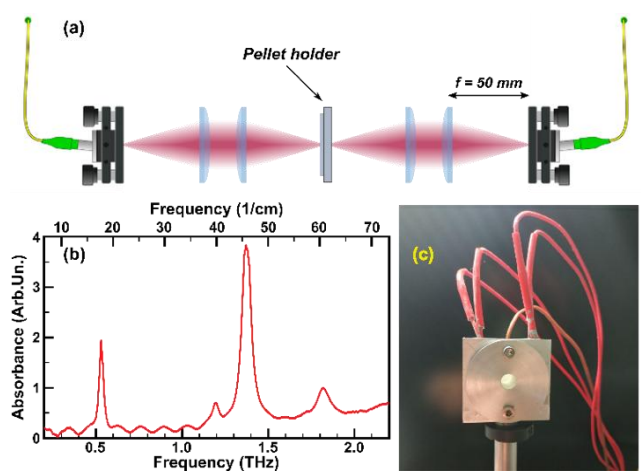

*Figure 1. Experimental setup: a) the optical scheme with the THz emitter and detector, four lenses (50-mm focal length) and the pellet holder with integrated heating. b) Typical THz absorption spectrum of α-Lactose monohydrate. c) Photo of the pellet holder with a pellet sample fixed inside, the red cables are used to power the heating cartridges, the orange one (thinner in the middle) is for the thermocouple.*

The detailed description of the mount's design, performances, calibration and experiment automation is discussed elsewhere[80], together with additional data analysis of the temporal and thermal dynamics of the lactose phase transformation.

### 2.3 Dehydration

A thermogravimetric analysis (Netzsch STA449F3 Jupiter apparatus) coupled with quadrupole mass spectrometry (Aëolos QMS403D, 70 eV, electron impact) via a heated capillary system has been set up to follow the dehydration process of the lactose samples during heating. We note that the performed analysis differs from a 'traditional' one where the temperature increases at a constant rate. Instead, the following procedure, similar to the TDS experiment, was followed: 1) increasing of the temperature to ~ 377 K with a constant rate of 5 K/min; 2) retaining of the sample at a constant temperature of 377 K for 80 minutes. Before starting the heating experiment, the thermogravimetric system was first evacuated and then flushed with ultrahigh purity helium. The experiments were carried out under dynamic inert gas atmosphere (helium: 99.999 purity) at a flow rate of 80 cm$^3$/min. The sample (4.5 mg) was heated in an $Al_2O_3$ crucible up to the desired temperature.

## 2.4 X-Ray Diffraction

The X-Ray Diffraction (XRD) analysis of lactose samples (of pellets grounded to powder) was done using a SmartLab Rigaku diffractometer equipped with a 9 kW rotating anode X-Ray generator (Cu Kα=1.5418 Å), in convergent beam vertical transmission geometry. The 2θ scans were performed in the 10-50° range, with a step size of 0.01° and a speed of 20°/min. The powder samples were inserted between two thin Mylar films.

## 2.5. Computational methods

### 2.5.1. Single molecule

The geometries of a lactose molecule and a lactose molecule plus a single water molecule were optimized in vacuum using the generalized gradient approximation as derived by Perdew, Burke, and Ernzerhof (PBE)[81] for the exchange and correlation functional, together with the Dunning correlation-consistent polarized quadruple-zeta (cc-pVQZ) basis set (PBEPBE/cc-pVQZ). Vibrational frequencies were subsequently calculated at the optimized geometries at the same PBEPBE/cc-pVQZ level. The empirical dispersion correction for density functionals (D3) with Becke and Johnson (BJ) damping[82] has been added during geometry optimization and frequency calculations. The rotational and translational modes have been projected out as standard procedure of the frequency calculations. The optimized structures presented no imaginary modes, indicative of a ground state geometry. Calculations have been performed with Gaussian 16[83].

### 2.5.2. Molecular crystals

First principles calculations were performed within the framework of Density Functional Theory (DFT) using the Projector Augmented Waves (PAW) method as implemented in the Vienna Ab initio Simulation Package (VASP) code[84,85]. As for the gas-phase molecules, the exchange and correlation functional was approximated using the generalised gradient approximation as derived by Perdew, Burke, and Ernzerhof (PBE)[81]. Within the context of a frozen core approximation, the 1s electrons were considered as core-electrons for C and O. As dispersive forces play a central role in the formation of molecular crystals, we included Van der Waals interactions through the DFT-D3 method improved with Becke-Johnson damping as devised by Grimme et al.[82]. The kinetic energy cut-off of the plane wave basis set was set to 600 eV.

The initial models of the different morphologies were obtained from various sources (as indicated in Table 1). Since no hydrogen positions can be derived from X-ray experiments, the hydrogen atoms in the initial models are placed based on an energy minimization procedure[86,88–91]. Before frequency calculations, the systems are fully relaxed (atomic positions and lattice parameters) in a geometry optimization. During structure optimization, atomic positions and cell parameters were optimized simultaneously. The equilibrium volume was determined using a Rose-Vinet based equation of state fit, as discussed elsewhere[50,92]. All structures were optimized using a conjugate gradient method with an energy-based convergence criterion of 1.0E-7. The resulting structures are well optimized with the final remaining forces acting on a single atom being below 1 meV/Å. The first Brillouin zone was sampled using an extended Γ-centred k-point grid (3×1×4 for α-Lactose, 4×2×2 for β-Lactose and 4×3×1 for αβ-Lactose), with the total energy of the system converged to < 1 meV for the entire system. The final crystallographic parameters of the different model systems are presented in the Supplementary Information (SI). For self-consistency purposes, reference calculations for gas-phase molecules were also performed under periodic boundary conditions, in a box containing at least 15 Å of vacuum in each direction. The first Brillouin zone was sampled using the Γ-point only.

Table 1. Source of the initial structures for the various models.

| System | Symbol | Waters per unit cell (one unit cell contains two lactose molecules) | Source of starting model | Ref. |
|---|---|---|---|---|
| α-Lactose Hygroscopic | α-LH | 0 | Experimental | 86 |
| α-Lactose Hygroscopic (A) | α-LHa | 1 | Manual addition of $H_2O$ to optimized structure of [86] at position given by [87] | 87 |
| α-Lactose Hygroscopic (B) | α-LHb | 1 | Manual removal of one $H_2O$ of α-Lactose Monohydrate from [88] | 88 |
| α-Lactose Monohydrate | α-LM | 2 | Experimental | 88 |
| α-Lactose Anhydrous | α-LA | 0 | Experimental | 89 |
| β-Lactose | β-L | 0 | Experimental | 90 |
| αβ-Lactose | αβ-L | 0 | Experimental | 91 |
| α-Lactose molecule | α-Lm | single molecule | Constructed from [86] | |
| β-Lactose molecule | β-Lm | single molecule | Manual rotation of hydroxyl group in optimized α-Lactose molecule | |

The vibrational spectra were calculated using the HIVE-program[93], developed by one of the authors. They were obtained through diagonalization of the Hessian matrix. In case of the molecular crystals, the translational modes are projected out; while for the free lactose molecules both translational and rotational modes were projected out[11,55,93]. The optimized structures presented no imaginary modes at the Brillouin-zone centre, indicative of high quality ground state geometries.

### 2.5.3. Modelling of the partial incorporation of water molecules into the molecular crystals

To study the role of water in the lactose molecular crystal lattice, we followed two routes. First, manual inclusion of a single water-molecule (*i.e.*, one water molecule per unit cell containing two lactose molecules) at roughly the position found in the literature[87] – α-Lactose Hygroscopic (A) (Table 1). Second, starting from the α-Lactose Monohydrate structure published by J.H. Smith[88], removing 1 of the 2 water molecules to obtain the same stoichiometry as before – α-Lactose Hygroscopic (B) (Table 1). The atomic structures of all optimized systems are included as *.cif* files in the SI. The crystallographic parameters of the optimized structures are presented in Section 5.2 of the SI.

## 3. Results and discussion

### 3.1. THz absorption properties with increasing temperature

The α-LM sample pellet was heated from room temperature to 377K. The absorption spectra at increasing temperatures are shown in Figure 2a. Four intense absorption peaks at around 0.53 THz, 1.17 THz, 1.37 THz and 1.80 THz are observed in good agreement with other reports by THz-TDS[54,64,94]. We observe a monotonous decrease and a "red-shift" of the peaks with the increasing temperature. Note also the signatures of the THz absorption by water vapour becoming visible at 377 K – an indication of water molecules being removed from the lactose crystals, starting at this temperature.

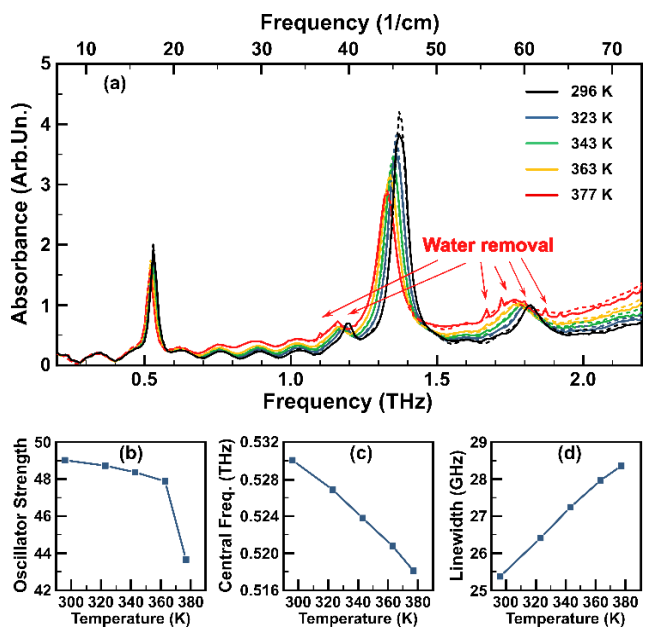

*Figure 2. a) Evolution of the THz absorption spectrum with temperature, solid lines – retrieved in the experiment, dashed lines – fit results. b)-d) Evolution of, respectively, the strength, central frequency and linewidth of an oscillator used to fit the experimental time-trace and corresponding to the characteristic α-Lactose monohydrate absorption line at ~ 0.53 THz.*

The origin of the spectral changes may be attributed to several mechanisms, such as lattice dilatation and thermal population of the phonon bands. To quantify the effect and – more importantly for the current discussion – to extract the peak central frequencies with a better precision, we have treated the experimental time-trace data with the Fit@TDS software[69]. This approach allows one to retrieve the material parameters and thus to more accurately interpret the results and their comparison to computational ones. We observe a reversible transition, but we don't go into further detail on these specific observations as these fall outside the scope of this manuscript and will be discussed elsewhere[80]. For now, it is sufficient to note that with temperature a) the peak width increases and b) there is an evolution of the oscillator parameters (*e.g.*, Figure 2b-d temperature evolution of the parameters of the oscillator corresponding to the ~ 0.53 THz absorption line). Note that the spectral parameters are identical during both heating and cooling ramp of the same sample, indicative for a reversible process.

### 3.2. Removal of water molecules. Phase transformation

At around 377 K, water removal starts, which is observed from the appearance of water vapour absorption lines - most visible in the 1.1-1.9 THz at 377 K (*cf.* the Figure 2a). The dehydration process was followed and the THz absorption spectra collected (Figure 3a). The effect of the removal of water from the lactose crystal lattice becomes clearly visible at temperatures slightly above 373 K (100°C). This process takes a certain amount of time depending on the temperature of the sample. In the ideal case, the temperature should be kept at ~100°C, but it would take several hours for the process to be completed. At a slightly elevated temperature, the process takes less time (*e.g.*, ~ 20 minutes at 125°C), however, this also increases the risk of uncontrolled mechanisms: such as the formation of different phases and other irregularities inside the sample. After a number of tests, we decided to keep the temperature at 377 K (104°C) low enough for a stable and controlled process, while at the same time high enough to complete the dehydration in a reasonable time. The spectral evolution at a fixed temperature of 377 K during 90 minutes shows visible water vapour absorption lines for about 30 minutes. After 60-70 minutes, no significant spectral change was observed, indicating a transformation to a more dehydrated crystal phase.

A thermogravimetric analysis allows the thermal behavior of lactose to be followed during the annealing process while at the same time tracking the gaseous species expelled from the sample. The resulting evolution of the sample mass loss is presented in Figure 3e. A mass loss is observed at around 373 K, with the weight stabilizing again after ~ 80 minutes at 377 K. The total mass loss during this process was ~ 1.8%, which is lower than the estimated initial water content of α-LM of 5-6%[79].

As noted in the previous section, the water removal around 377 K is consistent with a phase transition, giving – after cooling the sample back down to room temperature – rise to a spectrum that differs from the α-LM case presented in Section 3.1 (Figure 3d). This indeed indicates an irreversible process, consistent with the suggested phase transition. The THz absorbance spectrum of this phase looks similar to a phase reported earlier by Zeitler *et al*, which these authors refer to as α-Anhydrate[94]. For this new, second phase, the central frequencies of the most intense absorption peaks found at 23°C are the following: 0.88 THz, 1.125 THz, 1.22 THz, 1.39 THz, 1.545 THz.

In the spectra presented in Figure 3a, we observe that during the dehydration process no spectral shift of the absorption peaks occurs. We note however, a decrease in their intensity as well as the appearance of several new absorption peaks. The characteristic peak at 0.53 THz disappears during the dehydration process, while the 0.88 THz peak appears. After the time-trace data treatment, it is especially illustrative to consider the dynamics of the phase transformation process through the evolution of these two characteristic peaks. The evolution of the corresponding oscillators' central frequency and strength are shown in Figure 3b-c. This illustrates that the 0.88 THz peak is not present in the first phase, but appears in the second phase, while the 0.53 THz peak disappears completely in the second phase. However, the situation with the 1.37 THz peak is not trivial. Being previously widely attributed to α-LM, the 1.37 THz does not disappear completely during the dehydration process (i.e., stabilizes at ~ 50 % of its initial amplitude) and will be further discussed in section 3.4.4.

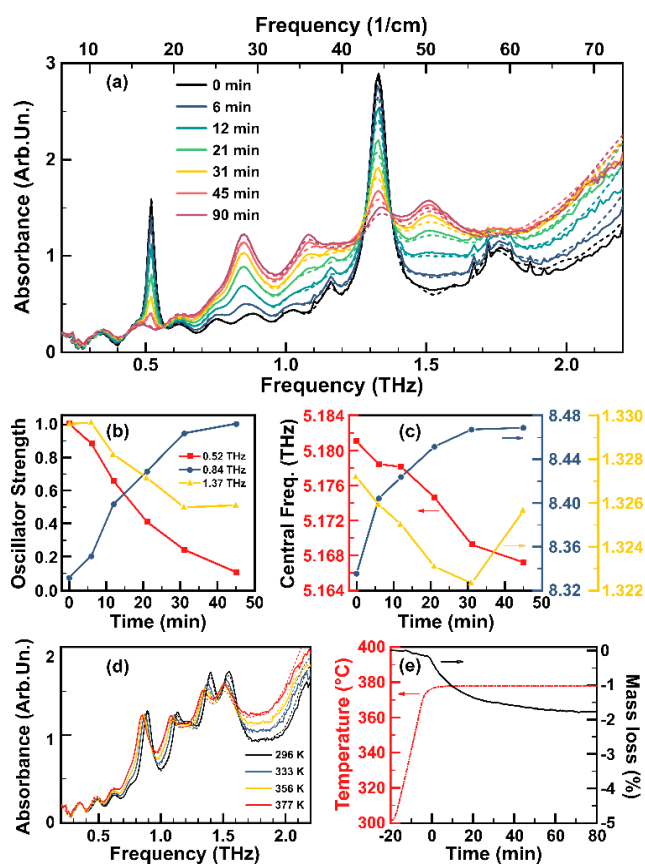

*Figure 3. a) Evolution of THz absorbance spectrum with time at the fixed temperature of 377 K, solid lines – retrieved in the experiment, dashed lines – fit results. b) Temporal evolution of the normalized strength of three oscillators used to fit the experimental time-traces, corresponding to two characteristic and well-distant THz absorption lines at ~ 0.53 THz (red) and at ~ 0.88 THz (blue), as well as the ~ 1.37 THz line (yellow). c) Temporal evolution of the central frequency of the three absorption lines. d) Evolution of the THz absorbance spectrum with the temperature decrease down to room temperature after 90 minutes at 377 K, solid lines – retrieved in the experiment, dashed lines – fit results. e) The repetition of the heating process in the mass-spectroscopy experiment.*

### 3.3. Further heating. Third phase

Upon further heating of the sample discussed above and increasing the temperature up to 500 K, using a procedure similar as before[i], we observe a second dramatic change in the spectrum at around 423 K (150°C): at 420 K the four spectral signatures present a significant decrease in intensity.

We therefore retained this temperature for around 20 minutes. At this point, any further spectral changes become insignificant, indicating the complete transition to a third phase. The transition to this phase is accompanied with further water loss, confirmed by the thermogravimetry experiments. Increasing the temperature further above 423 K (150°C), no other new spectral features appeared except during the lactose thermolysis at ~ 200°C. All experimental spectra of the three stable phases are summarized in Figure 4(a).

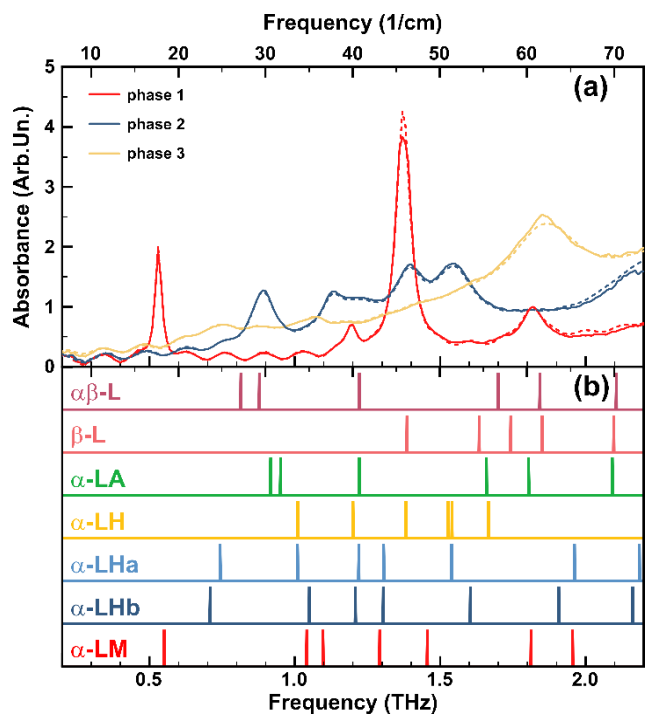

*Figure 4. a) The three observed lactose phases: THz absorbance spectra at the room temperature, solid lines – experimental results, dashed lines – corresponding fit. b) Calculated low THz vibrational modes at the Γ-point (colour vertical lines). In this figure, we have chosen to show all experimentally obtained and calculated spectra to introduce the discussion of Section 3.4.4., where the contributions of different lactose configurations to the three experimentally observed phases are discussed.*

### 3.4. Computational results and discussion

The experimental results provide a clear and consistent picture, with the most important conclusion that upon dehydration and accompanying phase transition, absorption peaks do not shift, but appear or disappear. Despite this clarity, the THz spectra alone are insufficient for the assignment of the specific lactose morphologies and based on the experimental results, a clear identification of the different phases was not possible. Although the specific modes are related to unique vibrational modes of the material, the frequencies themselves do not provide direct information on the atoms involved nor their configuration. This information is only available through indirect methods: by modelling the system and the associated vibrations and comparison of the resulting spectrum to that of the experimental observation. In this section, we will gradually build a theoretical model to describe the experimental observations discussed above. At each step, relevant insights and observations are highlighted and incorporated into the model at hand and insights about the nature of the different phases are provided.

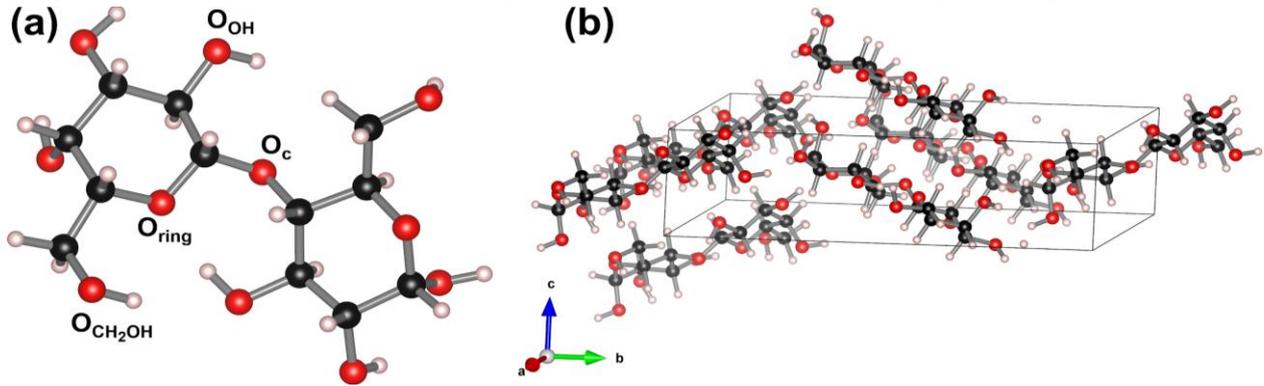

*Figure 5. Ball-and-stick representation of a gas-phase molecule α-Lactose (a) and the hygroscopic α-Lactose molecular crystal (b). The unit cell of the latter is indicated with a box. The different O sites are indicated in the gas-phase molecule.*

### 3.4.1. Pure α-Lactose in gas phase and molecular crystal

The simplest representation of α-lactose is the free gas-phase α-lactose molecule (*cf.*, Figure 5a). Although, for a single lactose molecule a wealth of different conformations exists, *e.g.* pyranose puckering conformations, we use the lowest energy sugar conformation, obtained starting from the geometry of α-lactose in molecular crystals. Once optimized, using standard DFT-based methods and functionals (see Computational methods, Section 2.5), the vibrational spectrum is easily obtained and the lowest vibrational modes are found at 0.70, 1.08, 1.26 and 1.83 THz (PBEPBE + D3/ cc-pVQZ). This clearly shows the THz frequencies considered a fingerprint of α-lactose (*i.e.*, 0.53, 1.19, and 1.37THz) to be missing, or at the least very poorly approximated, which is not solved by using different functionals or even different DFT implementations. For example, by calculating the molecular frequencies using plane-wave DFT (PBE+D3) and a periodic boundary condition representation of the gas phase molecule, the resulting lowest vibrations modes are 0.93, 1.27, 1.33, and 1.91 THz. Yet, although the calculated modes may give the impression that the discrepancy with experiment could be resolved by using a hypothetical perfect functional and algorithm, one should not be tempted into accepting the gas phase molecule as a suitable model for lactose molecular crystals[49].

As an initial crystalline model for lactose, we consider hygroscopic α-lactose (α-LH) under periodic boundary conditions (*cf.*, Figure 5b). We start from the experimentally obtained geometries presented in the literature, indicated in Table 1. As a result of the crystalline nature, the conformational freedom is severely restricted. Only the hydrogen atoms still retain some freedom, while the ring geometry (O—C torsion) is fixed (*e.g.*, no change in the pyranose puckering conformations, no transition between α and β form, etc.). However, even with this restriction, we note that since the PES is very flat, it allows for small local variation in the molecular geometry (O—C torsion, position of the H-atoms) (Appendix, B), which in turn give rise to slight shifts in the position of the THz peaks. Note that our theoretical model of α-LH contains no water molecules, which does not completely represent the actual situation, as experimental work pointed out that in anhydrous polymporhs, there is still some residual water present[79]. This choice for α-LH as initial model, compared to the experiments that were performed using α-LM, gives the opportunity to gradually build up the model, starting from molecules, over completely water-free crystals, to the final, fully hydrated α-LM. This will provide more clarity on the data. Starting from the atomic geometry published by Platteau *et al.*[86], the vibrational spectrum is obtained[92,95]. The presence of long ranged acoustic interactions, which are completely missed by single-molecule approaches[95], is indicated by the acoustic bands having an imaginary frequency in small regions of the Brillouin zone. Upon increasing the supercell size, these regions become vanishingly small[92]. The vibrational spectrum at the centre of the Brillouin zone (the Γ-point), however, shows no imaginary modes, indicative of a ground state geometry. Furthermore, due to the long wavelength of THz photons, only vibrational excitations at wave vectors q→Γ are experimentally observed, as the photons have a negligible momentum in comparison to the phonon momentum[96]. As such, the vibrational density of states (DOS) can be limited to the Γ-point only. Finally, as the molecular crystal contains two lactose molecules per unit cell, there is a doubling of vibrational modes compared to the gas-phase molecule configuration.

*Table 2. Calculated values of the lowest ten vibrational modes in THz (cm$^{-1}$). The different systems are indicated as in Table 1.*

| System | α-Lm | α-LH | α-LHa | α-LHb | α-LM | α-LA | β-L | αβ-L |
|---|---|---|---|---|---|---|---|---|
| Frequency, THz (cm$^{-1}$) | 0.697 (23.26) | 1.010 (33.70) | 0.743 (24.78) | 0.709 (23.65) | 0.551 (18.37) | 0.917 (30.60) | 1.386 (46.23) | 0.815 (27.19) |
| | 1.082 (36.08) | 1.201 (40.05) | 1.010 (33.68) | 1.050 (35.03) | 1.041 (34.73) | 0.951 (31.73) | 1.634 (54.51) | 0.878 (29.28) |
| | 1.261 (42.06) | 1.382 (46.11) | 1.221 (40.71) | 1.208 (40.30) | 1.097 (36.59) | 1.221 (40.74) | 1.742 (58.12) | 1.223 (40.78) |
| | 1.833 (61.13) | 1.526 (50.92) | 1.305 (43.54) | 1.304 (43.50) | 1.292 (43.09) | 1.659 (55.33) | 1.851 (61.74) | 1.699 (56.66) |
| | 2.224 (74.18) | 1.541 (51.39) | 1.539 (51.35) | 1.604 (53.49) | 1.456 (48.57) | 1.804 (60.18) | 2.097 (69.95) | 1.843 (61.48) |
| | 2.722 (90.78) | 1.666 (55.56) | 1.962 (65.44) | 1.908 (63.64) | 1.812 (60.45) | 2.091 (69.76) | 2.293 (76.48) | 2.105 (70.22) |
| | 2.940 (98.07) | 2.437 (81.30) | 2.185 (72.87) | 2.162 (72.12) | 1.955 (65.20) | 2.476 (82.58) | 2.461 (82.09) | 2.202 (73.46) |
| | 3.17 (105.74) | 2.666 (88.91) | 2.398 (79.98) | 2.510 (83.74) | 2.352 (78.45) | 2.585 (86.22) | 2.566 (85.61) | 2.401 (80.09) |
| | 3.77 (125.75) | 2.865 (95.58) | 2.622 (87.46) | 2.595 (86.55) | 2.585 (86.22) | 2.758 (91.97) | 2.612 (87.12) | 2.677 (89.30) |
| | 3.84 (128.09) | 3.058 (101.99) | 2.918 (97.35) | 2.921 (97.44) | 2.774 (92.52) | 2.929 (97.71) | 2.697 (89.96) | 3.007 (100.31) |

The calculated ten lowest vibrational modes of α-LH are shown in **Erreur ! Source du renvoi introuvable.**. The results show the lowest THz modes to be positioned at clearly higher frequencies, compared to molecular lactose α-LM.

This provides a good apparent agreement with the second (1.19 THz) and third (1.37 THz) mode observed in experiment. However, the experimental mode at 0.53 THz is missing entirely.

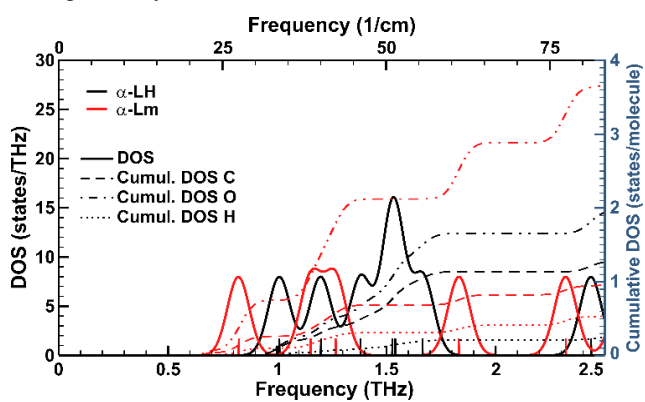

*Figure 6. The Γ-point vibrational DOS of the hygroscopic α-Lactose molecular crystal (black) and gas-phase α-Lactose molecule (red), as obtained from the periodic calculations. The exact positions of the vibrational modes are indicated with the vertical black and red lines. A Gaussian smearing of 0.05THz was employed to generate the DOS. The cumulative atom-projected DOS for the different atomic species is indicated by the black (crystal) / red (gas-molecule) dashed (C), dash-dotted (O), and dotted (H) curves, respectively.*

Comparison of the calculated molecular and crystal phonon DOS show a substantial change (*cf.*, Figure 6). Moreover, investigation of the atom-projected vibrational modes shows there is a strong reduction in vibrations of the O and H atoms of the system when going from a gas-phase molecule to a molecular crystal. The cumulative atom-projected phonon DOS per species shows this clearly in Figure 6. Although the O atoms are responsible for the bulk of the vibrational modes below 2 THz in both systems, the contribution is almost halved in the molecular crystal. In addition, the small vibrational contribution of the H atoms is halved going from gas-phase molecules to molecular crystals. In contrast, the contribution of the C atoms increases significantly. The lock-step trends of the O and H curves combined with their comparative relative reduction indicates the hydroxy groups are being restricted in their motion by the crystal structure, which is not unexpected considering the molecules of molecular crystals are bound via hydrogen bridges and dispersive interactions. This quenching of vibrational modes due to intermolecular forces is in line with the observations of Banks *et al.*[49]

This hypothesis is further confirmed by detailed investigation of the atom-projected vibrational spectra of the different O atoms in the system. The lowest four modes of the gas-phase molecule are governed by the two O atoms of the hydroxymethyl groups and six O atoms of the hydroxyl groups (*cf.*, Figure S1). Considering the same O atoms in the molecular crystal shows a significant reduction in their contribution to the vibrational spectrum. In contrast, the remaining three O atoms only provide a minor contribution to the low THz vibrational spectrum of the gas-phase molecule, which does not change much between the molecule and the molecular crystal.

### 3.4.2. Interaction with water molecules

*The hydroxyl and hydroxymethyl groups of the lactose molecule can move freely in the gas-phase configuration. The interaction with water-molecules as well as the molecular crystal configuration restrict the movement of these groups to varying extend. To investigate the impact on the vibrational spectrum of restricting the movement of the hydroxyl and hydroxymethyl groups, we consider both the gas-phase α-lactose molecule in interaction with gas-phase water molecules and the introduction of water molecules into the molecular crystal. Introducing one water molecule, placed at different positions around an α-lactose molecule (here, three energetically similar positions are considered and optimized to a ground state geometry, A, B, and C, see*

*Table 3, the compared systems contain exactly the same number of atoms; as such, the energy difference indicated in*

Table 3 is trivially the difference in total energy with position A set as reference) shows that adding a water molecule shifts the spectrum to lower frequencies, irrespective of its position. Water interacting through H-bonding with α-lactose (with a heavy-atom to heavy-atom O--O distance and O—H--O angle of 2.8 Å, 156°; 2.8 Å, 170° and 2.9 Å, 163° for position A, B and C respectively) can take either a donor or an acceptor role in the formed hydrogen bond. However, the donor-acceptor role taken by water does not give major changes to the spectrum (See Table S5 of the SI).

*Table 3. Gas-phase α-lactose molecule in interaction with gas-phase water molecule. Frequencies calculated at the PBEPBE+D3/cc-pVQZ level.*

| | Single α-lactose molecule | + $H_2O$ (position A) | + $H_2O$ (position B) | + $H_2O$ (position C) |
|---|---|---|---|---|
| System | | | | |
| ΔE, kcal/mol | – | 0.00 | -3.13 | 0.41 |
| Frequency, THz ($cm^{-1}$) | 0.697 (23.26) | 0.616 (20.54) | 0.586 (19.54) | 0.557 (18.59) |
| | 1.082 (36.08) | 1.032 (34.44) | 0.945 (31.51) | 0.677 (22.57) |
| | 1.261 (42.06) | 1.120 (37.35) | 1.178 (39.28) | 1.126 (37.55) |
| | 1.833 (61.13) | 1.217 (40.60) | 1.555 (51.88) | 1.234 (41.15) |
| | 2.224 (74.18) | 1.820 (60.72) | 2.116 (70.58) | 1.766 (58.92) |
| | 2.722 (90.78) | 2.230 (74.36) | 2.515 (83.90) | 2.041 (68.08) |

| 2.940 (97.98) | 2.676 (89.25) | 2.912 (97.15) | 2.320 (77.38) |

In the case of the molecular crystal, we incorporated a single water molecule using the two strategies mentioned in the computational methods section: 1) manual inclusion of a single water-molecule – α-LHa and 2) starting from the α-LM structure removing one of the two water molecules – α-LHb. The two configurations were optimized to similar, though distinctly different structures. The equilibrium volumes after structure optimization are found to be 747.15 and 746.20 Å$^3$ for the α-LHa and α-LHb, respectively, In these periodic structures, the water molecule interacts with four neighbouring lactose molecules through H-bridges (with a heavy-atom to heavy-atom O--O distance and --H--O angle of 2.9 Å, 155°; 2.7 Å, 177°; 2.8 Å, 172°; and 2.8 Å, 165° for the four interactions in the α-LHa system, and 2.9 Å, 151°; 2.7 Å, 178°; 2.7 Å, 176°; and 2.8 Å, 164° for the four interactions in the α-LHb system). The most significant difference between the two structures are the angles between the different lattice vectors (*cf.*, Table S1 of the SI). Of these two structures, the α-LHb structure is slightly lower in energy. However, this energy difference of a mere 23 meV per unit cell also indicates that the potential energy landscape (PES) is very shallow. As such, one may expect a sizable region of the PES to be sampled during room and elevated temperature experiments. The ten lowest vibrational modes are shown in **Erreur ! Source du renvoi introuvable.**. Comparison of these vibrational modes presents two interesting observations. First, the modes of the two configurations show only small variations, which may indicate that the basic properties of the system are sufficiently well described even if an experimental system may be moving on the PES (also see Figure S2 of the SI). Second, a vibrational mode has appeared at ~0.7 THz (~24 cm$^{-1}$) which is not present in the calculated anhydrous α-LH spectrum. Investigation of the atom-projected vibrational spectrum, shown in Figure 7, shows this mode is mainly due to the oxygen atoms of the lactose molecules (the blue curve). The oxygen atom of the water molecule (the red curve) also contributes to this mode, but not more than an average oxygen atom of the lactose molecule. Instead, a major contribution of the water-oxygen interaction is found in modes at about 2.6, 4.5, 5.0, and 6.4 THz (87, 150, 167, and 213 cm$^{-1}$). Further investigation of the atom-projected spectra of the different O functional groups (*cf.*, Figure 5a) in the lactose molecules, shows the O atoms of the hydroxyl and hydroxymethyl groups to contribute equally to the new mode at ~ 0.7 THz (~ 24 cm$^{-1}$). The remaining oxygen atoms of the lactose molecules only provide a marginal contribution. The presence of the water molecule appears to loosen the hydrogen bridges between the lactose molecules allowing a long ranged collective mode to appear. These findings are in line with the molecular gas phase spectrum, in which the lowest vibrational modes show the clear involvement of the water molecule in the motion.

The inclusion of a single water molecule in the lactose unit cell (2.6 % $H_2O$) could be considered as an intermediate step towards the formation of α-LM (5.0 % $H_2O$), and has slightly less water present than the actual water loss of 1.8% observed by the transition from the first α-LM to the second phase (which contains 3.2% water after the observed water loss; Section 3.2, Figure 3e). In experimentally prepared samples of hygroscopic α-lactose (and often indicated as being anhydrous), an initial water content of 1.34%[79] is found, which is about half of our theoretical structure with 2.6% . As such, based on the water content, the α-LHa/b might represent hygroscopic α-lactose more precise than our theoretical water free α-LH model. Comparison of the vibrational modes of these systems (α-LM and α-LHa/b) indicates no further additional low THz modes are created upon increasing water content; instead, the positions of the lowest modes are found at lower THz frequencies (*cf.*, **Erreur ! Source du renvoi introuvable.**).

Next, α-LM crystals were considered, using the experimental geometry of Smith *et al.*[88] as starting point. The calculated spectra show that the 0.53 THz (17.7 cm$^{-1}$) fingerprint mode of lactose is successfully recovered at a calculated frequency of 0.551 THz (18.4 cm$^{-1}$). Note that this result is obtained from DFT calculations on a perfect infinite periodic crystal at 0K. There is no need for a scale factor as is often used in literature[49,92]. This underlines the powerful nature of the used methods. Excellent agreement is obtained with the finite size experimental system, which probably contains defects and impurities, may present variations in the local water content and water molecule orientation, and is measured at elevated temperatures. The impact of the exact lattice volume and water orientation is considered in Appendix A, and shows variations of a few 0.01 THz on the calculated fingerprint frequency.

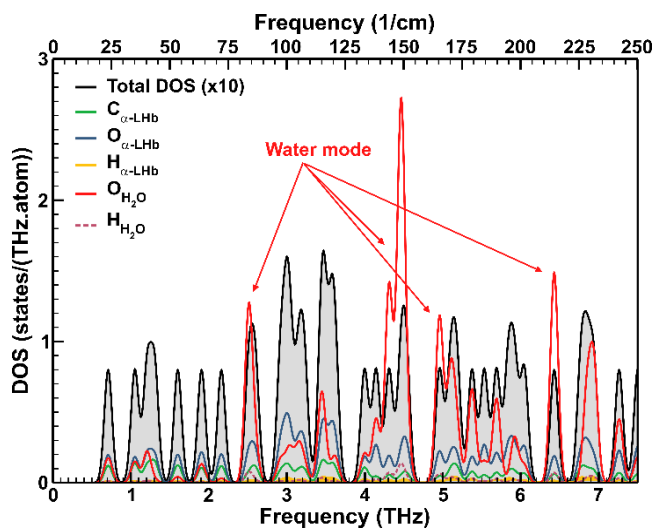

*Figure 7. Atom-projected vibrational spectrum at the Γ-point for the α-LHb configuration. The total DOS of the system is scaled by a factor 10 to improve comparison. Note that the α-LHb configuration contains two lactose molecules, accounting for 24, 22, and 44 atoms of C, O, and H, respectively. For the atoms belonging to the Lactose molecules, the spectrum is averaged per atom. For the atoms of the water molecule a similar averaging is performed.*

Investigation of the atom-projected vibrational spectra of α-LM presents a picture very similar to that of the systems with a single water molecule (*cf.*, Figure S3 of the SI). Again, it is interesting to note that the lowest vibrational mode shows mainly contributions of the O atoms of the hydroxyl and hydroxymethyl groups. So again, the water molecules, do not contribute directly to this mode, but indirectly by loosening up the molecular crystal structure. Each of these watermolecules form H-bridges with four neighbouring lactose molecules (with a heavy-atom to heavy-atom O--O distance and O--H--O angle of 2.9 Å, 147°; 2.7 Å, 177°; 2.7 Å, 175°; and 2.8 Å, 166° for one water molecule, and 2.9 Å, 147°; 2.7 Å, 177°; 2.7 Å, 175°; and 2.8 Å, 166° for the four interactions of the second water molecule). This stresses the importance of the use of periodic models for THz modelling of molecular crystals[49]. It also highlights the risk of incorrect mode assignment due to gas-phase modelling, as agreement of vibrational mode frequencies does not imply the presence of similar vibrations.

### 3.4.3. Formation energy of the different lactose morphologies

When trying to associate theoretical models to experimental systems, one generally considers the relative stability of the different theoretical models. In this work, we estimate the stability of the different molecular structures via their formation energy per unit cell ($E_f$) calculated as:

$$E_f = E_{cr} - \sum_i N_i E_i ,$$

with $E_{cr}$ the total energy of the optimized molecular crystal, $E_i$ the total energy of an α- or β-lactose, or water molecule in gas phase, and $N_i$ the number of these molecules present in the crystal unit cell. Note that we do not divide by the number of molecules as this would provide a misleading picture, since the number of water molecules differs in the different systems. However, since the number of lactose molecules is the same in each system (c.q. 2), $E_f$ presents the total binding of the molecular crystal as a whole, which is relevant for the crystal stability. For completeness, we also calculated the binding energy of the water molecules to the crystal, as the change in total binding energy compared to α-LH (cf., Table S1 of the SI). These latter values provide further insight in the stability of water in the crystal lattice.

Based on the existing literature, we have extended our set of lactose morphologies, being α-LM, α-LHa/b and α-LH with anhydrous α-Lactose (α-LA), β-Lactose (β-L), and αβ-Lactose (αβ-L)[89–91]. These structures were fully optimized (cf., Table S1 of the SI for the final lattice parameters) and their vibrational spectra were calculated (cf., **Erreur ! Source du renvoi introuvable.**). Remind here as well, that in all our constructed anhydrous models, no water molecules were included, following the experimentally proposed crystal structures[86–91]. This, however, may contradict the actual experimental situation[79].

The obtained formation energies, presented in Table S1, show that the incorporation of water significantly stabilizes the α-lactose phase, and more interestingly, the second water molecule appears to be bound slightly stronger than the first one. The α-LH is less stable than α-LA, which is in a good agreement with literature, showing the anhydrous α-lactose, to be slightly more stable than the hygroscopic α-lactose [97].

The slight increase in stability of the anhydrous β- and αβ-lactose compared to hygroscopic α-lactose is due to the slightly lower stability of the β-lactose molecule in the gas-phase (37 meV). In absolute terms, however, the β-L, αβ-L and α-LH configurations can be considered nearly degenerated in energy. These calculated formation and binding energies corroborate the experimental observation that it is energetically favourable for lactose to form molecular crystals, and that such molecular crystals tend to absorb water, which is consistent with the results of Listiohadi et al.[79], showing the absorption of water by the anhydrous polymorphs, even at low relative humidity during storage. It even indicates that a monohydrate (1 water/lactose, thus two waters per unit cell, containing 2 lactose molecules) is more stable per water, than a single water per unit cell (in the α-LHa/b models). Based on the calculated formation energies, the morphological path due to a dehydration, in a theoretical set up, is expected to be the following: α-LM (2 water molecules per unit cell) → α-LHa/b (1 water molecule per unit cell) → α-LH/α-LA/β-L/αβ-L (water-free). As the final anhydrous forms are relatively close in energy, the practical experimental synthesis path will play an important role in determining final product. This is in line with the generally accepted methodologies for producing different anhydrous lactose morphologies starting from α-LM[90,98–100].

### 3.4.4. Understanding the observed morphological changes from a computational perspective using calculated THz spectra

Using our calculated vibrational spectra, a direct comparison to the experimentally obtained results can be made (considering the discrepancies between a theoretical and experimental set up, c.q. water content). The different calculated spectra as well as the experimental spectra are shown together in in Figure 4. It rapidly becomes clear that reality may be somewhat more complex than the expected morphological path of α-LM → α-LHa/b → α-LH/α-LA/β-L/αβ-L.

Before we continue, it is important to remember two important aspects of calculated spectra. (1) As only modes which interact with the electromagnetic field of the THz beam give rise to an experimentally observed spectral feature, some of the calculated modes will be experimentally invisible due to symmetry and/or selection rules. (2) For our well-optimized theoretical structures – where, for example, Pulay stresses have been compensated – the calculated modes are at (almost) the exact experimental locations. The impact of water rotations and slight variations of the volume result in shifts of the order of several tens of GHz only (cf., Appendix). This removes the need for scaling of the calculated spectrum as is often done in literature[101–104]. As such, comparison to experiment means allowing for calculated peaks to be missing in experiment, but experimental peaks should not be missing in the calculated spectrum and should furthermore be almost perfect positional matches.

In Figure 8, we show the experimentally obtained spectra of the three phases in combination with the respective calculated spectra of the composing lactose configurations for each phase, proposed in this discussion. This allows for easier validation of the arguments of this section by the reader. IR-intensities were calculated for the different systems and presented for comparative purposes with other calculations in SI (Section 5.6.).

The starting α-LM phase of the sample, shown in Figure 8a, gives the low THz spectrum as obtained from the initial sample. It presents four distinct peaks at 0.53, 1.19, 1.37, and 1.82 THz (17.7, 39.7, 45.7, and 60.7 cm$^{-1}$), respectively, and is known in literature as α-LM[49,105].

Our calculated spectrum of α-LM presents eight vibrational modes below 2.5 THz (83.4 cm$^{-1}$). The 0.53 and 1.82 THz (17.7 and 60.7 cm$^{-1}$) modes are perfectly accounted for by the theoretical calculations on α-LM; however, according to the theoretically calculated spectra, the 1.82 THz mode can also be attributed to β-lactose or α-lactose anhydrous. The 1.19 and 1.37 THz (39.7 and 45.7 cm$^{-1}$) modes appear to be missing in the theoretically calculated spectrum of α-LM and need to be analysed further.

Looking at the calculated spectra of the various lactose crystal configurations considered in this work, the intense mode at 1.37 THz (45.7 cm$^{-1}$) does present a perfect match to both β-L and α-LH. Furthermore, the 1.37 THz (45.7 cm$^{-1}$) peak increases with 10-20% in intensity when recording the spectra during purging in N (compared to air), supporting the assignment of this peak as α-LH, under the assumption of dehydration during purging. Alternately, some calculated modes of α-LM also appear in the vicinity of the 1.19 and 1.37 THz (39.7 and 45.7 cm$^{-1}$) peaks. One might even argue the 1.37 THz (45.7 cm$^{-1}$) peak to be the result of the overlap of two modes at 1.29 and 1.46 THz (43.1 and 48.6 cm$^{-1}$). Some authors have assigned the calculated mode at 1.46 THz (48.57 cm$^{-1}$) to the experimental mode at 1.37 THz (45.7 cm$^{-1}$)[105]. Interestingly, in the full brillouin zone (in the y-direction, which is in the longest direction of the molecules), these two modes move toward one-another and cross at about 1.37 THz, at roughly 1/3 of the Γ-Y line. Therefore, this could be an indication these are contributing to the peak and an explanation for the relatively large discrepancy.

The low intensity mode at 1.19 THz (39.7 cm$^{-1}$), could also be attributed to several of the lactose crystal configurations: either the anhydrous α-LH, α-LA, or αβ-L or the partially hydrated α-LHa or α-LHb. This clearly indicates that a standard lactose monohydrate sample is a mixture of lactose configurations, of which α-LM presents the major fraction. This peak seems to be also present in the second phase, but not in the third one. It gives arguments to attribute this peak to α-LH or α-LHa/b. Bear in mind that complete water-free α-LH is not the experimental reality, as there is residual water in these polymorphs and as such, the experimental situation based on the water-content is something in-between the α-LH and α-LHa/b forms. At this point, it is interesting to note that commercially available α-LM still contains a small fraction of β-lactose molecules in the molecular crystal lattice (≤ 4% β-lactose).

Although β-L presents a calculated vibrational mode at 1.386 THz (46.23 cm$^{-1}$), a suitable candidate for the 1.37 THz mode, the strong observed intensity combined with the low concentration makes it unlikely to be the only contributing conformation. Interestingly, in experiments with α-LM / β-L mixed samples, by Yamauchi *et al.* it was shown that with increasing concentration of β-L, the 1.37 THz (45.7 cm$^{-1}$) reduces – but it does not vanish completely – while the 1.19 THz (39.7 cm$^{-1}$) increases in intensity[51]. Although the calculated spectrum of pure β-L does not show any peaks near 1.19 THz, αβ-L on the other hand does present this peak, which may indicate anomeric rotations to have occurred in the mixed samples reducing their purity.

Experimentally, a phase transition at 377K was observed (see Section 3.2.). The resulting second phase gives rise to the THz spectrum presented in Figure 8b. This transformation occurred simultaneously with the experimental observation of water vapour escaping out of the system. About 1.8% of mass loss has been recorded in the thermogravimetric experiment (Figure 3e). The new phase has a fingerprint of six modes, measured at 0.89, 1.13, 1.22, 1.39, 1.55 and 1.80 THz (29.7, 37.7, 40.7, 46.4, 51.7, and 60.0 cm$^{-1}$). Comparison to the simulated spectra shows that the visible α-LM mode at 0.53 THz (17.7 cm$^{-1}$) has disappeared, corroborating a phase transition away from the α-LM. As before, there are clear indications for the presence of a mixture of lactose configurations. The modes at 0.89 and 1.22 THz (29.7 and 40.7 cm$^{-1}$) may be due to either αβ-L or α-LA, of which the latter mode might also be an indication of α-LH, α-LHa, or α-LHb. Similar as for the first phase, the mode at 1.39 THz (46.4 cm$^{-1}$) can be assigned to either α-LH or β-L, or to a residual presence of α-LM in the sample, as clearly not all the water has been evaporated. The mode at 1.55 THz (51.7 cm$^{-1}$) could be due to either α-LH or α-LHa. Finally, the mode at 1.80 THz (60.0 cm$^{-1}$) can only be assigned to the α-LA. Interestingly, the mode at 1.13 THz (37.7 cm$^{-1}$) shows no good match with any of the considered lactose phases, and as such is an indication of the presence of another, possibly intermediate phase.

For this phase and as well for the starting one, we proposed a wealth of different lactose configurations. It is reasonable to assume that the number of configurations should be limited, therefore we continue our analysis of the first two experimental phases, of which the number of possible configurations could be reduced based on lattice strain arguments. It is reasonable to assume that the selected configurations contributing to the experimental α-LM phase and phase 2 should not give rise to obvious observable effects, we therefore expect the different phases to have an as similar as possible lattice structure, which would limit strain in the sample. Taking a closer look at the lattice parameters of the different lactose crystal configurations, given in Table S1 and in good agreement with the literature[89–91,100], we note that α-LA, αβ-L, and β-L have a unit cell volume which is about 7% smaller than α-LM. Furthermore, the latter two support lattice angles, which differ significantly from those of α-LM. Combining all the aspects, we propose that the α-LM the peak at 1.19 THz (39.7 cm$^{-1}$) can be accounted for by the α-LH, rather than to the αβ-L. In addition, the intermediates between α-LM and α-LH (*i.e.*, α-LHa or α-LHb) could naturally occur due local water gradients in the sample consisting of small crystals and as such possessing a very large surface region, and could also account for the 1.19 THz mode. The 1.37 THz peak can be accounted for by the α-LH, while the 0.53 and 1.82 THz peaks are perfectly accounted for by α-LM. We therefore suggest that our initial α-lactose monohydrate samples are mixtures of α-LM (core of the crystallites) and α-LH (the surface region of the crystallites), with a minimal contribution of beta-lactose. Also, for the second phase, we can reduce the number of possible contributing lactose configurations based on the secondary information of lattice strain. The αβ-L could be required for the measured modes at 0.89 and 1.22 THz (29.7 and 40.7 cm$^{-1}$). However, these modes are also covered by the α-LH and α-LA as well, to which the other four experimental modes are also assigned. As a result, we propose the second experimental phase to be a mixture of the α-LH and α-LA, having some residual water content, which is in agreement with what has been experimentally found for the preparation of these anhydrous configurations[79].

Finally, a second phase transition to the experimental phase 3 was observed in our sample at 420 K. This phase transition was accompanied by further water loss. The low THz spectrum is presented in the lower panel of of Figure 8. This phase presents only one obvious (and very broad) mode at 1.86 THz (62.0 cm$^{-1}$). Fitting of the experimental spectrum, however, indicates the presence/necessity of two additional modes at 0.736 and 0.986 THz (24.6 and 32.9 cm$^{-1}$). Comparison to the calculated spectra of the various phases shows that the clear mode at 1.86 THz (62.0 cm$^{-1}$) perfectly aligns with a mode of the αβ-L, β-L and α-LA spectrum. Each of these three modes is consistent with the water losses observed in the literature. Moreover, the anhydrous β-lactose is experimentally known to contain less water than anhydrous α-lactose[79]. This would indicate mutarotation between the α- and β- lactose anomers in the molecular crystal, which might not be unexpected at temperatures above 140°C[90]. Furthermore, the peak at 0.89 THz (29.7 cm$^{-1}$) of the second phase, assigned to α-LA has vanished. These observations may suggest the αβ-L and β-L configurations to be more probable than the α-LA of presenting a significant contribution to our third observed phase. There remains however the very strong base signal, indicating the formation of amorphous phase (see paper of McIntosh *et al.*[106]).

### 3.4.5. Identification of lactose configurations contributing to the three experimentally observed phases by the X-Ray diffraction

To further support our findings for the lactose configurations contributing to the three phases obtained in the experiment, we performed X-Ray Diffraction (XRD) of similar lactose samples.[1]

First, three samples have been prepared, reproducing the three stable phases discussed above. Three pellets were prepared from the α-Lactose Monohydrate powder using the same material, equipment and pellet parameters as discussed in Section 2.1. One pellet was kept at the room temperature, while the two others were prepeared on the heated capillary system of the TGA setup (Section 2.3.), reproducing the procedure of the THz-TDS experiment (*i.e.*, temperature increase from 296 K to 377 K at a rate of 3 K/min, followed by ~90 min at 377 K; and in case of the third pellet, a further temperature increase to 423 K, after which the third pellet was retained at 424 K for 20 min; finally all samples cool down to room temperature). Hereafter the pellets were studied in our THz-TDS setup to compare the spectra with the ones obtained in the experiment discussed above (Sections 3.1.-3.3., 3.4.4.). The results are presented in Figure 9a. The spectra of the first and the second phase are in a very good agreement with the ones observed in the first experiment (discussed in Section 3.1-3.3; Figure 4); we note however the small difference in the intensities of the THz peaks – this can be explained by the slight difference (~100 μm) in the thickness of the studied pellets. The difference seen in the spectra of the third phase, on the other hand, is due to the fact that we have used two different means to heat the samples (See Section 2.2.): the heating system of the TGA setup (used to prepare the samples for the XRD experiment) and the mount with integrated heating (used to prepare the samples for the experiments discussed in section 3.1-3.3 (Figure 4)). The first one allows more uniform heating of the whole pellet – which is important for the following XRD study, while our mount with integrated heating creates a certain temperature gradient (with the temperature being the highest at the pellet edge and decreasing toward the pellet centre). This gradient is negligible in the THz-TDS experiment, as in these experiments, one studies only a small area of the sample in the THz beam waist (d ~ 0.3 mm), below a certain temperature – which we see in the spectra of the second phase (heated up to 377 K), for instance. However, at higher temperature the sample becomes non-uniform even within the small area covered by the THz beam spot. Thus, the third phase can be considered more 'pure' in the sample prepared for the XRD experiment.

Next, pellets have been grounded to powder and studied in the XRD setup. The results are presented in Figure 9b-d for samples (phases) 1-3 respectively, along with the simulated spectra for the best matching lactose crystal phases, generated using the *Powder Diffraction Pattern* tool of VESTA[107]. It suggests the 1st sample to be mainly composed of α-lactose monohydrate, in a good agreement with literature[108,109] referring, for instance, to characteristic peaks of α-LM at 2θ =12.5° and 16.4°. However, one cannot exclude the presence of other crystal phases (*e.g.*, the work of Miao and Roos[109], they refer to the characteristic peak of β-L at 2θ = 10.4; the peaks at 2θ = 9°-20° are identified as mixtures of α-LM and anhydrous mixture of α-β lactose in different molar ratio's.). In our sample, we do not observe any peaks of β-L; however, some peaks (*i.e.*, the ones around 19° and 21°) may be attributed to α-LH, which is in correlation with our predictions based on the calculated vibrational spectra (*cf.*, section 3.4.4.).

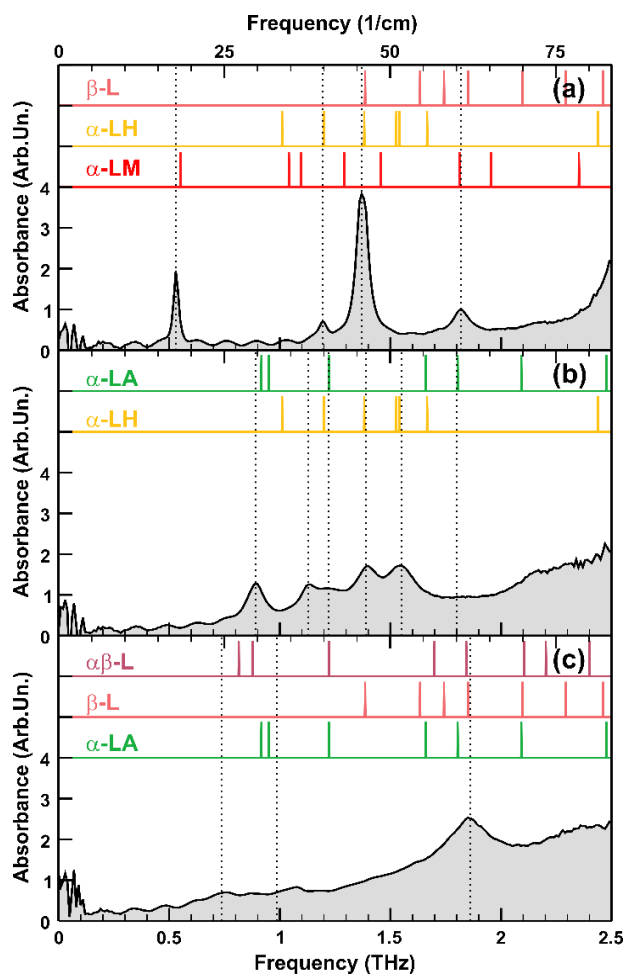

*Figure 8. Comparison of calculated low THz vibrational modes at the Γ-point (colour vertical lines) to spectra of the three experimentally obtained morphologies (black curves), (a)-(c) – phases 1-3, correspondingly. The vertical dotted lines indicate the position of the experimentally fitted modes.*

---

[1] *We note that the XRD experiments were performend during the revision process of our manuscript, thus it was not possible to use initial samples. However, we believe the results to be illustrative and deserve to be presented in the text of the paper.*

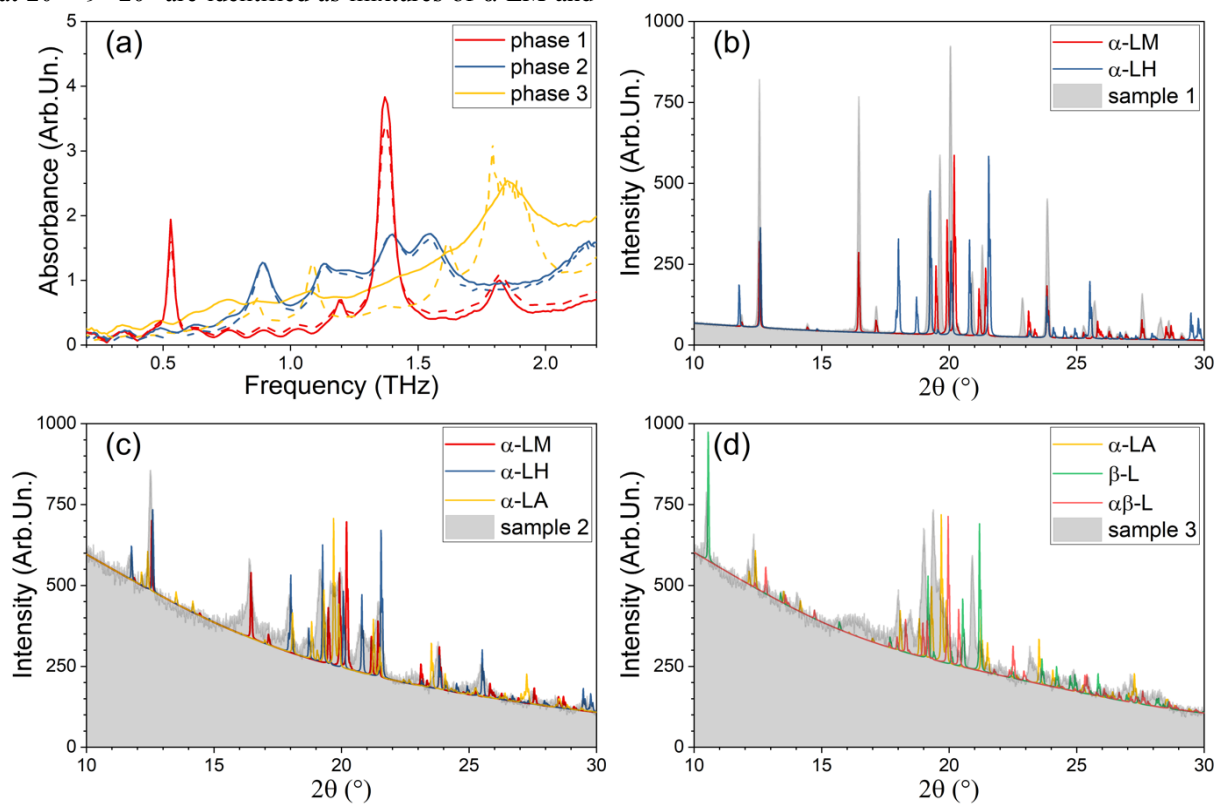

*Figure 9. a) THz absorbance spectra of the three phases at the room temperature: solid lines – the ones discussed in Section 3.1.-3.3. (pellets heated in the mount with integrated heating), dashed lines – the ones of the pellets prepared for the XRD experiment (heated in the TGA setup). b)–d) XRD data: grey curves - obtained for the studied samples (1-3, correspondingly), colour lines – simulated results for the corresponding lactose crystallographic phases.*

In sample 2, the situation appears to be more complicated: we do see the characteristic peaks of α-LH and α-LA as we suggested in section 3.4.4. (it also finds confirmation in literature[86]), however the characteristic peak at 2θ = 16.5° can be explained only by the presence of α-LM which is not inconceivable (it may be attributed to the re-hydration of lactose while grounding the pellet to powder). Moreover, it is not impossible that we were not able to detect it in the THz spectrum as the region around 1.1-1.6 THz is 'crowded' by multiple peaks (with the 1.1, 1.29, 1.46 THz and even 1.04 THz DFT-predicted peaks of α-LM could be presented in the region).

For sample 3, we see the peak of β-L at 2θ = 10.4° emerging in the spectrum, while the characteristic peaks of α-LM at 2θ = 12.5° and 16.4° are vanishing. The α-LA peaks remain, and the appearance of αβ-L is possible. This is in agreement with the interpretation given in section 3.4.4 based on calculated vibrational spectra and correlates with literature[89,110]. However, we note again that sample 3, discussed here, could be considerably different from the "phase 3" of Section 3.3. – it is clear from the comparison of their THz spectra (Figure 9a).

## 4. Conclusions

To summarize, we present a detailed study of the THz spectrum of one of the most used and thought-to-be-simple materials by the THz community, namely alpha-lactose monohydrate. In our experiments, our samples go through two phase-transitions. In our study, the THz spectrum of alpha-lactose monohydrate is investigated both experimentally (monitoring of the THz spectral evolution during heating of the lactose sample – reported for the first time) and theoretically (quantum mechanical periodic calculations on molecular crystals). By integrating our results, we come to the clear conclusion that the alpha-lactose monohydrate system is more complex than generally thought.

Firstly, we show that the gas phase single molecule approach for spectral modelling is insufficient for the THz range, as this concerns long ranged vibrational modes. Although the necessity for periodic calculations is not a new conclusion within the context of solid-state spectroscopy, we feel that underlining this aspect is still beneficial, especially in the experimental THz community. As such, this work aims to further stress the importance of good and accurate periodic DFT calculations. In contrast to the mid infrared spectroscopic range where the vibrations are localized around one or several molecular bonds, in the THz range, the vibrational modes are delocalized among several molecules. They form what one can call "phonons of the molecular crystal", making the use of periodic calculations inevitable, as was also pointed out by Banks *et al.*[49]. The clear corollary of this conclusion is that different crystal structures will give rise to different THz spectra. One can exploit this property experimentally and use THz spectroscopy to probe and discriminate the crystalline structure of a molecular crystal sample.

Secondly, we discuss the role played by the hydration water molecules in the THz vibrational spectrum. Our spectra show drastic changes with decreasing water content. From experimental work, these spectral changes could be attributed to the appearance/disappearance of the absorption peaks upon hydration. Although the water molecules have a larger freedom of motion than the carbohydrates, one cannot separate the vibrations of the carbohydrate lattice from the vibrations of the water molecules. Consequently, these water molecules and associated hydrogen bonds do not only play a perturbative role, but also participate in the formation of the vibration. More importantly, the water molecules' impact is not related to their donor-acceptor role, instead their presence loosens the hydrogen bridges between the lactose molecules allowing long ranged collective modes to emerge. They can be seen as having a lubricating role, allowing the motion of several rigid pieces together.

Thirdly, we show that the simple $X \rightarrow Y \rightarrow Z$ phase transformation picture is hard to maintain and moreover highly dependend on the sample preparation, as is clear from section 3.4.5. It appears the seemingly distinct phases and transformations in reality involve mixtures and/or additional intermediate phases. For example, the alpha-lactose monohydrate sample – widely used in the community for test purposes – cannot be explained by only one single crystal phase. In this work, we provide a first step by proposing a credible explanation for the well-known spectrum. However, further investigation of the crystal structure of alpha-lactose monohydrate will be needed to close the question in all its intricacies.

Finally, we also highlight the unfortunate nomenclature used for lactose phases. Specifically the identification as anhydrous phases, even though these are known to support a significant water (and variable) content[79]. The published crystal structures of the so-called anhydrous phases show no water molecules, pointing to incoherencies with the experimentally observed water content[79], and complicating theoretical corroboration of experimental results for such systems. These crystal structures thus require further work, which will be the result of combining different techniques, including X-ray diffraction analysis, crystal simulation and THz spectroscopy.

Overall, this work shows that THz frequency vibrational spectroscopy is a very challenging, yet very powerful tool both experimentally and theoretically. Even though THz spectroscopy exists already more than two decades, we believe that we are still experiencing its infancy. Further work and combination with other experimental and theoretical techniques will be needed to bring it to its full potential and to allow for its straightforward application on more complex structures such as protein crystals or even protein structures like amyloids or viral capsids.

# Appendix. Peak shifts and broadening

In our graphical representation of the calculated vibrational spectra, we have employed a Gaussian distribution with a variance of 0.05 THz (~ 1.67 cm$^{-1}$). As these are arbitrary choices, they are not indented to provide an exact representation of an experimental spectrum. Instead, they are introduced to provide some intuition on expected variations of intensity as well as peak broadening due to (almost) degenerate vibrational modes. The realities of experimental peak shapes are extremely complex, and beyond the scope of what is currently possible at the first principles DFT level. However, it is possible to gain some intuition about some contributing factors. In this appendix, we briefly discuss two factors that influence peak shapes: A. crystal lattice volume and B. water orientation. For both aspects, we will focus on the α-LM system.

## A. Crystal lattice volume

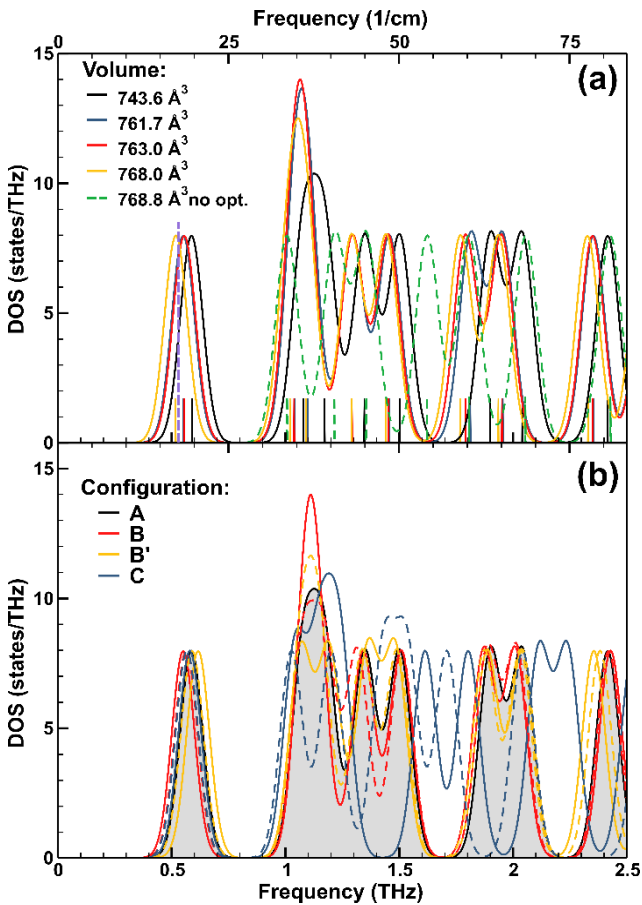

*Figure 10. Low THz vibrational modes at the Γ-point for the α-LM system. (a) Spectra calculated for optimized atomic geometries at varying unit cell volume. The experimental location of the α-LM fingerprint is indicated by the vertical dashed line. (b) Spectra calculated for configurations with different water orientations. Solid lines present the vibrational spectrum of relaxed configurations, while the dashed lines are obtained for the non-relaxed configuration.*

The low THz spectrum at five different unit cell volumes is presented in Figure a. The selected volumes represent (1) the equilibrium volume at the DFT level as a result of Pulay stresses: 743.6 Å$^3$, (2) the equilibrium volume at the DFT level corrected for Pulay stresses: 761.7 Å$^3$, (3) the experimental equilibrium volume: 768.8 Å$^3$, (4) and (5) optimized structures obtained as part of an energy-volume curve with volumes close to the corrected DFT volume (763.0 Å$^3$) and experimental volume (768.0 Å$^3$), respectively[88,92]. The structure (3) at the experimental equilibrium volume is obtained by scaling the DFT equilibrium structure at 743.6 Å$^3$ to the experimental volume without subsequent structure optimization at fixed volume, indicated as 'no opt.' in Figure a. The other structures are obtained the same way but with subsequent structure optimization at fixed volume.

The lattice parameters of the different structures are presented in Table S2 of the SI, as well as the relative total energy with regard to the minimal energy structure. These results show no radical structural transformations occur. Furthermore, for the optimized systems, the relative total energy variations are extremely small (less than 1 k$_B$T at room temperature). The impact of structure optimization on the total energy for the system at the experimental equilibrium volume is about 0.4 eV per unit cell. This stresses the need for further optimization after scaling the geometry of a molecular crystal[92].

Having a closer look at the low THz spectra clearly shows that scaling the structure and performing phonon calculations results in a rather distorted spectrum with the vibrational modes showing rather extreme shifts to both higher and lower frequencies of up to 0.5 THz (~ 16.7 cm$^{-1}$). Focussing on the optimized structures instead, a more consistent picture emerges. With increasing lattice volume, the modes shift to lower frequencies. However, the size of this shift appears to vary between modes.

If we direct our attention to the 0.53 THz (17.7 cm$^{-1}$) fingerprint mode (indicated by the purple dashed line), we clearly see it is in agreement with the calculated values for all four optimized structures. Note, however, that there is no linear relation between the lattice volume and mode frequency, as can be seen from comparing this mode for the 768.0 and 743.6 Å$^3$ systems to the 761.7 and 763.0 Å$^3$ systems. Based on this observation, we suggest that small local variation of the lattice parameters, induced by internal and external stresses, give rise to experimentally observed broadening of the vibrational modes. Furthermore, as can be seen in Figure a, this broadening will be different for each mode. Extending on this observation one could imagine different local defects or external stresses to give rise to a broadening of specific modes, providing access to a means of identifying such defects. This is, however, beyond the scope of the current work.

## B. Water orientation

The water molecules in the α-LM system are bound to the α-Lactose molecules through hydrogen bridges. Such bounds are much weaker than ionic or covalent bonds, and as such allow the water molecules to move relatively freely. To obtain a crude picture of the PES of the rotation of a single water molecule in the α-LM system, we sampled the PES created by the three rotation angles of the water molecule (Figure S4a of the SI). The total energy in each point is calculated using single-point calculations[ii]. The resulting PES is shown in Figure S5b, and indicates only a single minimum is present (the additional minima are symmetry-equivalent structures). We selected two configurations (B and B') of our sample grid close to the PES minimum (A), as well as a configuration (C) close to the PES energy maximum. For these three configurations, the vibrational spectrum was calculated, for both the non-relaxed as well as the relaxed geometries. This structure relaxation – during which the water molecules are kept frozen and the full lactose molecules are allowed to relax – leads to a small improvement of the relative energy of configuration B and B', while a significant improvement of the relative energy (3765 meV down to 453 meV) for the C configuration is observed (*cf.*, Table S3 of the SI). Furthermore, it is interesting to note that the lattice parameters do not differ significantly between the various configurations. There is, however, a consistent volume increase of the unit cell (1.5-5 Å$^3$) as a result of the rotated water molecule. Considering the presence of two water molecules per unit cell, the impact of their rotation thus could be a source of additional volume, resolving the last few Å$^3$ of volume discrepancy between the experimental and quantum mechanically modelled system.

The lowest vibrational modes are shown in Figure b and their positions are tabulated in Table S4. The impact of structure relaxation is limited to a few tens of GHz for the configurations close to the PES minimum. For configuration C, close to the PES maximum, the modifications are more pronounced, as is to be expected, since the non-relaxed structure gave rise to five imaginary modes, while only a single imaginary mode remained after relaxation. As such, configuration C presents a transition state configuration. It is interesting to note that, in spite of all these differences between the configurations, each configuration gives rise to the 0.53 THz (17.7 cm$^{-1}$) fingerprint mode, albeit at a slightly different frequency. Thus, the orientation of water has only a limited impact. This is also observed in the gas-phase molecule calculations, showing the minimal impact of the donor/acceptor role of water in the H-bond to lactose on the spectral modes (See Table S5 of the SI). This clearly shows that the free rotation of water does not alter the spectrum, but plays an important role in the peak broadening of the low THz spectrum, even for modes that are not directly assigned to the water molecules.

## Acknowledgements

We thank the referees for their review – it has allowed us to improve tha manuscript, and especially the excellent suggestion to check our findings by X-Ray diffraction. We thank Dr. Florent Blanchard and Dr. Pascal Roussel for their kind help with that experiment.

This work was partially supported by: i) the international chair of excellence "ThOTroV" from region "Hauts-de-France"; ii) the welcome talent grant "NeFiStoV" from European metropole of Lille; iii) the "TeraStoVe" grant from I-site ULNE; iv) the French government through the National Research Agency (ANR) under program PIA EQUIPEX LEAF ANR-11-EQPX-0025 and ExCELSiOR ANR 11-EQPX-0015; and v) the French RENATECH network on micro and nanotechnologies.

The computational resources and services used in this work were provided by the Centre de Ressources Informatiques (CRI) and by the VSC (Flemisch Supercomputer Center), funded by the Research Foundation Flanders (FWO) and the Flemisch Government – department EWI.

## Notes and references

[i] More specifically, using a new sample, we repeated the above experiment up to 377K. Then, after stabilization of the THz spectrum at 377K, we increased the temperature further but this time at a rate of 3K/min, as the heating unit was not powerful enough to sustain a 5K/min rate at temperatures in the range of 450-500K.

[ii] Our coarse sampling of the PES gave rise to 1900 grid points. To keep the computational cost in check (now 2 years of CPU-time), no structure optimizations are performed as this would increase the computational cost with a factor of the order 500. Note that such structure optimizations would still suffer from Pulay stresses, resolution of which would increase the computational cost another order of magnitude.